\documentclass[prl,twocolumn,showpacs]{revtex4}
\usepackage{graphicx,epsf}
\begin{document}

\def\tende#1{\,\vtop{\ialign{##\crcr\rightarrowfill\crcr
\noalign{\kern-1pt\nointerlineskip}
\hskip3.pt${\scriptstyle #1}$\hskip3.pt\crcr}}\,}

\def\a{\alpha}
\def\e{\varepsilon}
\def\d{\delta}
\def\t{\tau}
\def\n{\nu}
\def\o{\omega}
\def\s{\sigma}
\def\G{\Gamma}
 
\def\ra{\rightarrow}
\def\Ra{\Rightarrow}
\def\pd{\partial}
\def\bq{{\bf q}}
\def\om{\omega_m}
\def\omt{\tilde\omega_m}
\def\ot{\tilde\omega}
\def\ob{\bar\omega}
\def\ou{\omega_U}
\def\DD{{\cal D}}
\def\LL{{\cal L}}

\def\be{\begin{equation}}\def\ee{\end{equation}} 
\def\bea{\begin{eqnarray}}\def\eea{\end{eqnarray}} 
\def\nn{\nonumber} 
\def\lb{\label} 
\def\pref#1{(\ref{#1})}

\title{Signature of stripe pinning in optical conductivity}
\author{L.\ Benfatto and C.\ Morais Smith }
\affiliation{D\'epartement de Physique, Universit\'e de Fribourg, 
P\'erolles, CH-1700 Fribourg, Switzerland}

\begin{abstract}
The response of charge stripes to an external electric 
field applied perpendicular to the stripe direction is studied within a
diagrammatic approach for both weak and strong pinning by random
impurities. The sound-like mode of the stripes described as elastic strings
moves to finite frequency due to impurity pinning. By calculating the 
optical conductivity we determine this characteristic energy scale for
both a single stripe and an array of interacting stripes. 
The results explain the anomalous far-infrared peak observed recently 
in optical-conductivity measurements on cuprates.
\end{abstract}
\maketitle

Recent optical conductivity measurements in high-$T_c$ cuprates
have revealed the existence of a strong peak in the far-infrared
regime, which has been attributed to the presence of charge stripes
\cite{cu1,paolo}. The typical energy of this peak is much lower than
the mid-infrared band detected in previous experiments
\cite{uchida} and reproduced by numerical simulations of the
Hubbard model \cite{lorenzana}. A similar feature reported 
for slightly overdoped Bi$_2$Sr$_2$CuO$_6$ \cite{paolo2} was
explained in terms of a charge-ordering instability of the Fermi liquid
\cite{caprara}. However, these arguments cannot be extended to the
underdoped regime, where the Fermi-liquid description breaks down and the
charge degrees of freedom are known to order \cite{tranquada}.

In this paper we propose that this anomalous peak may originate from the
pinning of the zero-energy phonon-like mode associated with transverse
fluctuations of stripes, which shifts to finite energies in the presence of
impurities. We calculate the optical conductivity of striped systems using
a diagrammatic approach, which allows us to determine the peak frequency
for both weak and strong impurity pinning.  In the single-stripe case, we
find a peak at a frequency $\n$ given by the ratio between the stripe
velocity $v$ and a characteristic impurity length scale $\lambda$, which is
determined by the strength of the pinning.  For a stripe array, a second
energy scale $\ou$, proportional to the strength $U$ of the stripe-stripe
interaction, becomes relevant. In the weak-pinning regime, which is
appropriate for describing cuprates, a second peak appears around $\ou$. When
$\ou \sim \n$, a resonance arises and the main peak splits.  Our
calculations yield a good estimate for the peak frequency measured in La
cuprates \cite{cu1,paolo,lobo} and clarify some peculiar features observed
in experiments, such as the peak-sharpening \cite{cu1} and splitting
\cite{paolo} which occur when stripe-stripe interactions are important.

We begin by considering a single line of holes (stripe) embedded in a
two-dimensional $\ell\times L$ antiferromagnetic background with lattice 
constant $a$. The transversal dimension $\ell$ correspond to the 
interstripe distance when the stripe array is explicitly considered (see
below). 
The stripe is oriented along the $y$-axis and acts as a domain-wall for the
underlying magnetization. 
The holes are assumed to move in the transverse $x$
direction and their dynamics is governed by the $t-J$ model. In the
long-wavelength limit a wave-like action describes the transverse 
displacement $u(y)$ of the stripe with
respect to the equilibrium position \cite{prb98}. The stripe velocity 
$v=\sqrt{Jt}a$ is related to the kinetic energy of the holes ($t$) and the
staggered magnetization ($J$) of the surrounding regions. 
In terms of the dimensionless
displacement field $\psi = u/a$, the action reads
\begin{equation}
{\cal S}=\frac{1}{2at} \int d\tau dy \left[ \LL_0[\psi]+V[\psi] \right],
\label{sint}
\end{equation}
where $\LL_0[\psi]=(\partial \psi/\partial\tau)^2+v^2 (\partial \psi/
\partial y)^2$ and $V [\psi]$ is the pinning potential.  In analogy with
the standard problem of pinning of elastic manifolds
\cite{giamarchi,cdw1,cdw2}, the interaction of the (striped) density of
carriers $\rho({\bf r})$ with the short-range disorder potential $U({\bf
r})=U_0\sum_i\delta({\bf r}-{\bf r}_i)$ can be written as $\int d{\bf r}
U({\bf r})\rho({\bf r}) \sim \rho_0U_0\sum_i \cos[2\pi (u-x_i)/\ell]$,
where ${\bf r}_i= (x_i,y_i)$ denotes the position of the defects, and
$\rho_0$ is the average stripe density \cite{giamarchi}. We consider that
only defects localized near the stripe effectively act to pin it, so that
the cosine term can be expanded around the minimum to yeld, within the
quadratic approximation, $V [\psi] = V_0 at \sum_i (\psi(y)-\beta_i)^2
\delta(y-y_i)$. The impurities pin the local displacement $\psi(y_i)$ at a
value $\beta_i$, with positive strength $V_0$, and we assume $\hbar=k_B=1$.

In the absence of dynamical fluctuations, the stripe
accommodates the impurity potential by adopting a configuration
$\psi_0(y)$, which is a solution of the static equation of motion.  The
full, time-dependent solution $\psi(y,\t)$ is described as a fluctuation
around the equilibrium configuration $\psi_0(y)$, {\it i.e.},
$\psi(y,\t)=\psi_0(y)+\phi(y,\tau)$.  Introducing the Fourier transform
$\phi(q,\om)$ with respect to the momentum $q$ and the Matsubara frequency
$\om=2\pi m T$, the reciprocal-space action for the $\phi$ field becomes
\bea
{\cal S}=\frac{1}{2}\sum_{q,\om}\DD_0^{-1}(q,\om)
|\phi(q,\om)|^2 \nn \\
+\frac{V_0}{2}\sum_{q,q',\om}\phi^{\dagger}(q-q',\om)\phi(q,\om)S(q), 
\lb{sintf} 
\eea
where $S(q)=(1/L)\sum_i e^{iqy_i}$ is the form factor, 
and $\DD_0(q,\om)= at(\om^2+v^2q^2)^{-1}$ is the bare Green function.  

An electric field applied in the $x$-direction, perpendicular to the
stripe, generates a current $J=\int dy e(\pd \psi/\pd t)$. The linear
charge density of the stripe is denoted by $e/a$. From the Kubo formula for
the optical conductivity \cite{mahan}, $\sigma(\o)=e^2 i\omega
\DD(q=0,i\om\ra \o-i\d)$ and by analytic continuation in the lower
half-plane of the dressed Green function $\DD(q,\om)$ one finds
\be
\Re \s(\o)=-e^2 \o \Im \DD(q=0,i\om\ra \o-i\d).
\lb{defs} 
\ee
In the absence of impurities, $\DD$ coincides with the bare Green function
$\DD_0$. Its replacement in Eq.\ \pref{defs} yields the expected response
of a massless, sound-like mode, $\Re\s(\o)=e^2 at\d(\o)$.  The dressed
Green function $\DD$ may be evaluated from the Dyson equation,
$\DD^{-1}(q,\om)=\DD_0^{-1}(q,\om)-\Sigma=(\om^2+v^2 q^2-\Gamma t)/(at)$,
where we have performed a rescaling $\Sigma=\Gamma/a$ such that $\Gamma$
has dimensions of energy.  The self-energy $\Gamma$ is evaluated by
averaging the product of the several $S(q)$ factors, which appear in each
term of the perturbative series for $\Gamma$, over the random impurity
positions \cite{mahan}. Crossing diagrams, arising from scattering
processes by different impurities, correspond to increasing powers of the
impurity density $n_i$. If $\Gamma$ is evaluated up to order $V_0^2$,
crossing terms do not appear (Born approximation) and one obtains
\be
\Gamma(\om)=-\frac{V_0}{2}n_ia+\left(\frac{V_0}{2}\right)^2n_i a \frac{1}{L}
\sum_q D_0(q,\om),
\lb{wc}
\ee
where $\DD_0$ was used for calculating the second-order correction. On defining
the frequency $\o_0=n_i v=n_ia\sqrt{Jt}$, the dimensionless
quantities $\ot=\o/\o_0$, $G=\Gamma t/\o_0^2$, and the parameter $\a=V_0/
(2n_i aJ)$,  Eq.\ \pref{wc} takes the form
\be
G= -\a-\frac{i\a^2}{2\ot}.
\lb{sewcnsc}
\ee
According to Eq.\ \pref{defs} the optical conductivity is given by 
\be
\Re\s(\o)=
-\s_0\frac{\ot G''}{(\ot^2+G')^2+G''^2},
\lb{wcsc}
\ee
where $\s_0=e^2 at/{\o_0}$ and $G = G' + i G''$.  Note that as soon as a
real part is generated in $G$, the delta function in the optical conductivity
moves to a finite frequency and its amplitude is determined by $G''$. Using 
\pref{sewcnsc}, we find 
\be
\Re\s(\o)=\s_0\frac{2(\o/\n)^2}{4(\o/\n)^2[(\o/\n)^2-1]^2+\a},
\lb{wcnsc}
\ee
with $\n=\sqrt{\a}\o_0$. We note that $\Re\s(\o)\ra 0$ both as $\o\ra 0$
and as $\o\ra \infty$, and has a peak at $\o\simeq \n$, as shown by the
solid line in Fig.\ 1a.  The parameter $\a$, which controls the perturbative
expansion for $G$, depends both on the ratio $V_0/J$, {\it i.e.} the
competition between impurity potential and elastic energy, and on the number
of impurities per site, $n_ia$. Thus the weak-pinning limit,
where only first powers of $\a$ need be considered (as in Eq.\
\pref{sewcnsc}), is realized for a large number ($n_ia\sim 1$) of weak
($V_0/J\ll1$) impurity centers. In this case, Eq.\ \pref{wcnsc} shows that
the peak frequency scales according to $\n\simeq \sqrt{V_0 t n_i a}$,
moving toward zero energy as the strength of the pinning potential
decreases. In the evaluation of Eq.\ \pref{wc}, we used the bare Green
function $\DD_0$. The divergence of the self-energy \pref{sewcnsc} at small
$\o$ indicates the break down of this crude approximation at low
frequencies. Inclusion of the next-order term in the perturbative expansion
shows that higher-order diagrams become important for $\o<\sqrt{\a}\n/2$,
{\it i.e.} at frequencies much smaller than $\n$.  It has been argued
\cite{cdw1} that a better result for $G$ may be achieved by using the
dressed Green function in Eq.\ \pref{sewcnsc}, which leads to the
self-consistent equation $G=-\a+(\a^2/2)(-\ot^2-G)^{-1/2}$. However, the
resulting expression for $\s(\o)$ shows a spurious gap below $\o\approx 0.7
\n$, which is thought to originate from the fact that the self-consistent
Born approximation selects arbitrarily a subset of diagrams of higher order
in $\a$ without including crossing terms, whose contribution cannot be
excluded if $n_ia$ is large.  Because we are interested primarily in the
evaluation of the peak frequency $\n$, and not on details of the
low-frequency behavior of $\s(\o)$ \cite{fogler}, 
we restrict our analysis to the
non-self-consistent Born approximation, which provides a reliable estimate
both of the maximum of $\s(\o)$ and of its scaling with $\a$.

A different approach must be considered in the strong-coupling limit $\a\gg
1$, which corresponds to a small number ($n_ia\ll1$) of strong
($V_0/J\gg1$) pinning centers. In this case one might expect that the entire
perturbative series in $\a$ should be resummed, but there exists a
criterion for selecting the relevant higher-order diagrams. For $n_ia\ll
1$, it is possible to
select the subset of diagrams of first order in $n_i$ after averaging
over the impurity distribution ($T$-matrix approximation) \cite{mahan}. By
explicit summation of the resulting series, using the full Green function $\DD$
while evaluating each diagram, one obtains a self-consistent equation for
$G$. In the limit $\a\ra \infty$, one finds the analytical solution 
$\Re\s(\o)=2\s_0\sqrt{\ot^2-1}/\ot^3$ for $\ot>1$ and $\Re\s(\o)=0$ for
$\ot<1$ \cite{cdw2}. Thus, $\Re\s(\o)$ displays a gap at $\o_0$ and a
maximum at a frequency $\o$ slightly above this value.
When the solution is evaluated at
arbitrary $\a\gg1$, one finds essentially the same features. In
this case the requirement of full self-consistency is crucial, and the gap
feature arises from the strong pinning of the sound mode. The difference
between the weak- and strong-pinning results can be understood by
considering that in both cases the peak frequency corresponds to the
sound mode of a free string, which is now trapped on a characteristic
length scale $\lambda$. When few strong impurities are present, $\lambda$
is given by the average distance between impurities $1/n_i$ and the peak
appears 
at the frequency $\o_0$. However, when the number of impurities increases
and $V_0$ decreases, the elastic energy cost of accommodating the string on a
length scale $1/n_i$ is too high, and $\psi$ is pinned on the Larking length 
$\lambda_{loc}$ \cite{giamarchi,larkin}.
Dimensional estimates based on
Eq.\ \pref{sint} yield $\lambda_{loc}\sim
v/\sqrt{V_0n_iat}$, which indeed corresponds to the characteristic
frequency $\n$ in the weak-pinning regime.
\begin{figure}[hbt]
\begin{center}
\includegraphics[width=8.5cm]{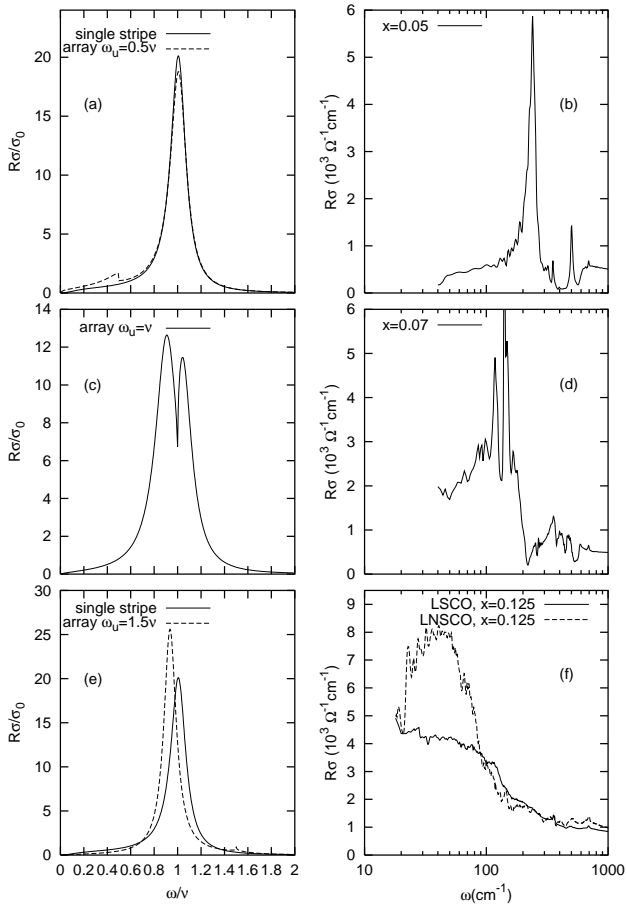}
\end{center}
\caption{ Left: real part of the optical conductivity, in units of $\s_0$, as a
function of $\o/\n$ evaluated in the Born limit ($\a=0.1$). 
When interaction between stripes is not relevant ($\ou<\n$, Fig.\ 1a) one finds
a single peak (the second one for the stripe array is almost suppressed), 
as observed in LSCO at $x=0.05$, Fig.\ 1b \cite{paolo}. When the
resonance condition $\ou=\n$ is satisfied, the peak at $\o=\n$ splits into
two equally-weighted peaks (Fig.\ 1c), as observed experimentally at $x=0.07$,
Fig.\ 1d \cite{paolo}. 
By increasing the inter-stripe interaction ($\ou=1.5\n$, Fig.\ 1e) 
the peak sharpens and moves to slightly lower
frequencies, as observed experimentally by comparing Nd-free with Nd-doped
LSCO compounds, Fig.\ 1f \cite{cu1}. For comparison with our
zero-temperature calculations, we display the experimental data at the 
lowest measured temperatures above $T_c$. Notice the logarithmic scale on
the right panels.}

\end{figure}

Until now we have concentrated on the transverse pinning of a single
stripe. However, a more realistic description of 
cuprates requires the study of an array of stripes. By considering a harmonic
coupling strength $U$ between neighboring stripes, which are labeled 
by an index $n=1, \dots N$, we obtain the action \cite{hass99}
$$
S=\frac{1}{2at}\sum_n\int d\tau dy \left\{ \LL[\psi_n]
+Ut(\psi_n-\psi_{n+1})^2+V[\psi_n]\right\}.
$$
Here $\psi_n=\psi_n(y,\t)$ and 
the fluctuating field $\phi_n(y,\t)\equiv \phi(x=n\ell,y,\t)$ is a function
of the discrete $x$ values. By repeating the same steps as before, we
observe that Eq.\ \pref{sintf} is still valid, on taking
$q\ra \bq=(q_x,q_y)$, $S(\bq)=(LN)^{-1}\sum_{ni}e^{iq_xn\ell+iq_yy_i^n}$, where
$y_i^n$ is the impurity coordinate at the $n$-th stripe, and $\DD_0$
becomes 
\be 
\DD_0(\bq,\om)=\frac{at}{\om^2+v^2q_y^2+4Ut\sin^2(q_x\ell/2)}.  
\lb{greenint} 
\ee 
The system of interacting stripes displays an anisotropic sound-like
mode at long wavelengths, with an effective elastic coefficient $U/a$ in
the $x$ direction, and an optical mode at $q_x \ell =\pi$ with energy
$\ou=2 \sqrt{Ut}$.

The general procedure used above to address the effect of
disorder is extended readily to the case of interacting stripes. We 
assume that the average density of impurities does not depend
on the stripe index $n$, {\it i.e.} $n_i^n=n_i$. In the Born limit, the main
difference with respect to the single-stripe case arises from the
additional integration in $q_x$ of the Green function
$\DD_0$ in Eq.\ \pref{wc}.  By introducing the rescaled variables and
$\ob=\o/\ou$, one obtains
\be
G=-a+\frac{\a^2}{\pi}\frac{\o_0}{\ou}\int_{0}^{\pi/2} dx 
\frac{1}{\sqrt{\sin^2x-\ob^2}}.
\lb{gsa}
\ee
When $\ob>1$ the second term in Eq.\ \pref{gsa}  contributes only to
$G''$. In the limit of vanishing stripe-stripe interaction $U\rightarrow
0$, or $\ob\ra \infty$, it reproduces the single-stripe result Eq.\
\pref{sewcnsc}.  For finite
$U$, a second energy scale $\ou$ is relevant for determining the optical
conductivity, which is defined by Eq.\ \pref{defs} with $q$ replaced
by ${\bf q}$.  The new feature is the divergence of $G''$ at $\ou$, which
leads to a second peak around this frequency, where $\Re\s(\o)$ vanishes. 
The vanishing of the conductivity at $\ou$ in the
weak-pinning regime should indeed be expected: at this energy, neighboring
stripes fluctuate in antiphase, and because the total current arises from
the sum of the currents of each stripe, the net current in such a case is
zero.
At $\ou\ll\n$ or $\ou\gg\n$ the peak at $\ou$ is almost completely suppressed
by the overall decrease of $\Re\s(\o)$ at small and high
frequencies, see dashed lines in Figs.\ 1a and 1e. 
However, when $\ou\simeq \n$ the peak at $\o\simeq \n$ splits (Fig.\ 1c), 
signaling a resonance between the characteristic
mode of the stripe array at the zone boundary and the characteristic
frequency of sound-like oscillation of each stripe at the Larkin
length.
Dimensional estimates indicate that this resonance arises when 
the average stripe separation coincides with the Larkin length.
Moreover, for the stripe array we observe a sharpening of the peak
for $\ou>\n$ due to the flattening of $G''$ around $\n$ (dashed
line in Fig.\ 1e). The softening of the peak frequency for $\ou>\n$ scales
according to $\a^2 \n/\ou$, so it never moves far away from $\n$.

In the strong-pinning regime the qualitative behavior of $\Re\s(\o)$ for
the stripe array is similar to that of the single stripe, and no resonance
is observed. Because the interaction with the impurities is the predominant
effect, the behavior of $G$ is almost entirely dominated by the pinning of
the sound mode at $\o_0$. More details about the strong-pinning regime will
be presented elsewhere \cite{unpu}.

We next compare our results with the available experimental data.  The
velocity of the unpinned sound mode may 
be estimated by using standard values $t \sim J \sim 0.1$ eV for cuprates.
The other relevant quantity is $n_i$. In these materials, the
chemical-substituted dopants, whose two-dimensional density
is  $x/a^2$ ($x$ being the doping),  simultaneously provide charge
carriers in the plane and act as pinning centers. As we explained at the
beginning, we consider that on average
the impurities within
a range $\ell$ contribute to pin each stripe, {\it i.e.} $n_i=x \ell/a^2$.  
Neutron diffraction 
measurements have shown that 
in the underdoped regime ($0.05<x<0.12$) 
of La$_{2-x}$Sr$_x$CuO$_4$ (LSCO) $\ell = a/2x$ \cite{tranquada}, 
which yields $n_i a = 1/2$ and $\o_0 \sim 0.05$ eV.  Because $n_i a\sim 1$
corresponds to the weak-pinning regime, we expect that the peak frequency
is given by $\nu=\sqrt{\a} \o_0$ and thus is reduced with respect
to the estimate of $\o_0$, in agreement with experimental data, see Fig.\
1b, 1d, 1f \cite{cu1,paolo}. As shown in Fig.\ 1b-1f, the peak frequency
softens as $x$ increases. Even though we
also expect a general decreasing of $\nu$ due to a screening effect on $V_0$,
a more quantitative comparison with experimental data would require the
inclusion of lattice effects, which is beyond the scope of this paper. 


Another interesting feature can be explained by our calculations: with
increasing doping concentration, stripe-stripe interactions become more
important and one should consider the stripe array.  Our results predict
that the peak splits when the average stripe separation $\ell$ is of the
order of the Larkin length, see Fig.\ 1c.  This splitting has been
observed: for $x = 0.05$ doped LSCO, a single peak appears (Fig.\ 1b)
whereas for $x = 0.07$ the peak splits (Fig.\ 1d) \cite{paolo}. Although
the value of $\a$ as a function of doping $x$ is in general not known, we
may nevertheless estimate the peak frequency where the splitting arises as
$\n = v / \ell = 0.014$ eV $= 113$ cm$^{-1}$, in excellent agreement with
the measured value (Fig.\ 1d) \cite{paolo}.  Our calculations also allow us
to understand the results obtained by Dumm {\it et al.} \cite{cu1} in
Nd-doped LSCO (Fig.\ 1f): the broad peak observed in LSCO \cite{cu1} moves
to slightly lower frequency and becomes sharper in the Nd-doped compound,
where the stripe array is ordered \cite{tranquada}. We have shown that when
inter-stripe interactions increase, the peak becomes slightly softer,
narrower and more intense (Fig.\ 1e), as observed experimentally.  Note
that in Ref.~\cite{cu1} the feature seen in the Nd-free sample
in Fig.\ 1f was interpreted as an anomalous large Drude peak.
 However, we claim that a stripe signature is still present in
these data, even though a partial overlap with the single-particle response
is possible.  Indeed, resistivity data from Ref.~\cite{cu1} give
$\s(0)\simeq 8 \cdot 10^3$ $\Omega^{-1}$cm$^{-1}$, which is much larger
than the value obtained by extrapolating the supposed Drude peak. Moreover,
a peak at nearly the same frequency 
is also seen at $x=0.12$ in Ref.~\cite{paolo}.

In addition, our results provide the correct order of magnitude for the peak
height. The one-dimensional optical-conductivity unit $\s_0\equiv
\s_0^{1d}$ converts into a three-dimensional one as $\s_0^{3d}=
\s_0^{1d}/\ell d =(2e^2/\hbar)(t/\o_0)(x/d)$, where $d$ is the interlayer
spacing. At $x=0.07$, using $d\simeq 6 $ \AA, we estimate the maximum of
$\s(\o)$ to be of order $10^3 \ \Omega^{-1}$cm$^{-1}$, as experimentally
observed \cite{paolo}.


Note that longitudinal transport, which is possible for half-filled
stripes as in the cuprates, was neglected in our approach. However, the 
experimental observation of the same far-infrared peak in nickelates
\cite{ni}, which exhibit filled stripes, supports our hypothesis that the 
anomalous peak in the far-infrared regime originates from transverse 
fluctuations. In fact, our results could be used also to describe the
nickelates, but in this case the effective $t$ and $J$ 
parameters within a one-band model are not known.

In conclusion, we calculated the contribution from stripe pinning to 
the optical conductivity in cuprates within a diagrammatic 
approach accounting for the presence of randomly distributed impurities. 
We found a peak in the far-infrared range at a frequency which has the 
same order of magnitude as observed in the experiments. In addition, our
results explain the splitting of the peak, as well as the narrowing
and increased intensity, which arise when stripe-stripe interactions are
relevant. 

We acknowledge fruitful discussions with D.\ Baeriswyl,
P.\ Calvani, S.\ Caprara, T.\ Giamarchi, J.\ Lorenzana, B.\ Normand and
J.\ Tranquada. We thank the authors of Refs.\ \cite{cu1,paolo} for providing
us with the experimental data.
This work was supported by the Swiss NSF under grant No.~620-62868.00.



\begin{thebibliography}{99}

\bibitem{cu1}  M.\ Dumm, D.\ N.\ Basov,
S.\ Komiya, Y.\ Abe, and Y. Ando, 
Phys.\ Rev.\ Lett.\ {\bf 88}, 147003 (2002).

\bibitem{paolo} A.\ Lucarelli, S.\ Lupi, M.\ Ortolani, P.\ Calvani, 
P.\ Maselli, M.\ Capizzi, P.\ Giura, H.\ Eisaki, N.\ Kikugawa, T.\ Fujita,
M.\ Fujita, and K.\ Yamada,
Phys.\ Rev.\ Lett.\ {\bf 90}, 037002 (2003).

\bibitem{uchida} S.\ Uchida, T.\ Ido, H.\ Takagi, T.\ Arima, Y.\ Tokura, and
S.\ Tajima, Phys.\ Rev.\ B {\bf 43}, 7942 (1991).

\bibitem{lorenzana} J.\ Lorenzana and G.\ Seibold, 
Phys.\ Rev.\ Lett.\ {\bf 90}, 066404 (2003).

\bibitem{paolo2} S.\ Lupi, P.\ Calvani, M.\ Capizzi, and P.\ Roy, 
Phys.\ Rev.\ B {\bf 62}, 12418 (2000).

\bibitem{caprara} S.\ Caprara, C.\ Di Castro, S.\ Fratini, and M.\ Grilli,
Phys.\ Rev.\ Lett.\ {\bf 88}, 147001 (2002).

\bibitem{tranquada} J.\ M.\ Tranquada, B.\ J.\ Sternlieb, J.\ D.\ Axe, Y.\
Nakamura, and S.\ Uchida, Nature {\bf 375}, 561 
(1995).

\bibitem{lobo} R.\ P.\ S.\ M.\ Lobo, F.\ Gervais, and S.\ B.\ Oseroff, 
Europhys. Lett. {\bf 37}, 341 (1997).

\bibitem{prb98} C.\ Morais Smith, Yu. A. Dimashko, N.\ Hasselmann, and A.\ O.\
Caldeira, Phys.\ Rev.\ B {\bf 58}, 453 (1998).


\bibitem{giamarchi} 
See T.\ Giamarchi and E.\ Orignac, {\em Theoretical Methods for Strongly 
Correlated Electrons}, Eds. D. Senechal {\em et al.}, Springer, New York,
2003, and references therein.


\bibitem{cdw1} H.\ Fukuyama, J.\ Phys.\ Soc.\ Jap.\ {\bf 41}, 513 (1976).

\bibitem{cdw2} H.\ Fukuyama and P.\ A.\ Lee, 
Phys.\ Rev.\ B {\bf 17}, 535 (1978).

\bibitem{mahan} G.\ D.\ Mahan, {\em Many-Particle Physics}, Kluwer Academic,
New York (2000). 

\bibitem{fogler} M.\ M.\ Fogler, Phys.\ Rev.\ Lett.\ {\bf 88},
186402 (2002).

\bibitem{larkin} A.\ I.\ Larkin and Y.\ N.\ Ovchinnikov, J.\ Low
  Temp.\ Phys.\ {\bf 34}, 409 (1979).


\bibitem{hass99} N.\ Hasselmann, 
A.\ H.\ Castro Neto, C.\ Morais Smith, and Y.\ Dimashko, 
Phys.\ Rev.\ Lett.\ {\bf 82}, 2135 (1999).


\bibitem{unpu} L.\ Benfatto, and C.\ Morais Smith, unpublished.

\bibitem{ni}  N.\ Poirot-Reveau, P.\ Odier, P.\ Simon, and F.\ Gervais,
Phys. Rev. B {\bf 65}, 094503 (2002).





\end{thebibliography}
\end{document}